# Model Spectra of the First Potentially Habitable Super-Earth - Gl581d

Lisa Kaltenegger
Harvard Smithsonian Center for Astrophysics, 60 Garden st., 02138 MA Cambridge, USA
also MPIA, Koenigstuhl 17, 69117 Heidelberg, Germany
lkaltene@cfa.harvard.edu

Antígona Segura
Instituto de Ciencias Nucleares, Universidad Nacional Autónoma de México, México

Subhanjoy Mohanty
Imperial College London, 1010 Blackett Lab., Prince Consort Road, London SW7 2AZ, UK

**Abstract**

Gl581d has a minimum mass of 7 $M_{Earth}$ and is the first detected potentially habitable rocky Super-Earth. Our models confirm that a habitable atmosphere can exist on Gl581d. We derive spectroscopic features for atmospheres, assuming an Earth-like composition for this planet, from high oxygen atmosphere analogous to Earth's to high $CO_2$ atmospheres with and without biotic oxygen concentrations. We find that a minimum $CO_2$ partial pressure of about 7 bar, in an atmosphere with a total surface pressure of 7.6 bar, are needed to maintain a mean surface temperature above freezing on Gl581d. We model transmission and emergent synthetic spectra from 0.4μm to 40μm and show where indicators of biological activities in such a planet's atmosphere could be observed by future ground- and space-based telescopes. The model we present here only represents one possible nature – an Earth-like composition - of a planet like Gl581d in a wide parameter space. Future observations of atmospheric features can be used to examine if our concept of habitability and its dependence on the carbonate-silicate cycle is correct, and assess whether Gl581d is indeed a habitable super-Earth.

**Subject headings**: astrobiology, Planets and satellites: atmospheres, composition, detection, individual: Gl581d, Earth, Instrumentation: spectrographs

## 1. INTRODUCTION

Spurred by the recent large number of radial velocity detections (Mayor et al. 2009) and the discovery of the transiting system CoRoT-7b (Leger et al. 2009), the study of planets orbiting nearby stars has now entered an era of characterizing massive terrestrial planets (aka super-Earths). The Gl581 system is a particularly striking example among these new discoveries, consisting of an M3V star orbited by a minimum of 4 planets, 3 of which are in the Earth to super-Earth range (Udry et al. 2007, Mayor et al. 2009). Of key interest are the two super-Earths located on either edge of the Habitable Zone (HZ) in this system. Detailed calculations (Selsis et al. 2007, von Bloh et al. 2007, Hu & Ding 2011) show that the inner super-Earth, Gl581c, is too close to its star to maintain water on its surface, and would enter a runaway greenhouse regime if no specific extreme cloud coverage can be invoked. It would thus loose all its water to space, leaving the planet hot and dry. Gl581d, on the other hand, was initially thought to lie on the outer edge of the HZ and could under some conditions be potentially habitable (see Selsis et al. 2007); the updated orbital period of 66.8 days (Mayor et al. 2009) places it within the Habitable Zone of the system with a semi-major axis of 0.22 AU. Its high eccentricity of 0.38 ± 0.09 brings it even closer to its host star. The mean flux received by a planet with an eccentric orbit is larger than that received on a circular orbit with the same semi-major axis (Williams & Pollard 2002, Spiegel et al. 2009, 2010; Dressing et al. 2010): the mean flux at Gl581d is equivalent to that for a circular orbit around Gl581 with $a = 0.20$AU (henceforth, unless specifically noted otherwise, we refer to this value as the orbital radius of Gl581d). Low mass Main Sequence M dwarfs are the most abundant stars in the galaxy, representing about 75% of the total stellar population. Many planets like Gl581d are thus likely to be found in the near future, providing excellent targets for future space missions (Scalo et al. 2007, Tarter et al. 2007, Kaltenegger et al. 2010b) searching for signposts of biological activity in planetary atmospheres.

The goals of our paper are to explore the atmospheric conditions under which this planet may be habitable (see also Hu & Ding 2011, Wordsworth et al. 2010, von Paris et al. 2010, von Bloh et al. 2007, Selsis et al. 2007) and the remote detectability of such spectral features and biosignatures (see e.g. Kaltenegger et al. 2010a, 2010b, DesMaraise et al. 2002). For these Earth-like assumptions, we investigate: a) the minimal atmospheric conditions for Gl581d to be potentially habitable at its current position, b) if habitability could be remotely detected in its spectra in transmission and through direct imaging and c) compare the resulting spectra to Earth's spectra and discuss differences in the chemistry due to the host star spectrum. Note that the model we present here only represents one possible nature of a planet like Gl581d in a wide parameter space that includes Mini-Neptunes. Without spectroscopic measurements we will not be able to break the degeneracy of mass and radius measurements (see e.g. Adams et al. 2008) and characterize a planetary environment.

Transmission and emergent spectra of terrestrial exoplanets may be obtained in the near future with the same techniques that have successfully provided spectra of Earth (see e.g. Christensen et al. 1997, Cowan et al. 2010, Irioni 2002) and extrasolar giant planets (EGP) (see e.g. Grillmair et al. 2008, Swain et al. 2008). Emergent spectra of rocky planets in the HZ are dominated by reflected starlight in the visible to near-IR and thermal emission from the planet in the mid-infrared, while transmission spectra result from starlight that is filtered through the planet's atmosphere. Such spectroscopy provides molecular band strengths of multiple transitions (in absorption or emission) of a few abundant molecules in the planetary atmosphere. We generate synthetic spectra of Gl581d from 0.4 μm to 40μm to explore which indicators of biological activities in the planet's atmosphere may be observed by future ground- and space-based telescopes such as the Extremely Large Telescope (E-ELT) and the James Webb Space Telescope (JWST).

The organization of this paper is as follows. We introduce the host star Gl581 in §2, outline our spectroscopic and atmospheric models in §3, and present our results – atmospheric gas and temperature profiles, implications for habitability, spectroscopic signatures and planet-to-star contrast ratio – in §4. In §5 we discuss some important additional uncertainties such as the influence of $CO_2$ clouds and the concentration of other greenhouse gases, and summarize our conclusions in §6.

## 2. THE STAR GL581

Gl581 is an M3 dwarf at a distance of 6.3 pc from the Sun, with absolute magnitudes in the V and K bands of $M_V = 11.56 \pm 0.03$ and $M_K = 6.86 \pm 0.04$ respectively (Bonfils et al. 2005b, hereafter B05b, and references therein). From the V-band bolometric correction derived by Delfosse et al. (1998; BCV = -2.08), B05b then infer a bolometric luminosity $L_{Star}$ = 0.013 $L_{Sun}$. From the mass-MK relationship (Delfosse et al. 2000), B05b also find a mass $M_{Star}$ = 0.31 $M_{Sun}$. We adopt these values unchanged. For this mass, the solar-metallicity evolutionary tracks (Baraffe et al. 1998; specifically, the tracks denoted BCAH98 models.[1]) imply a radius $R_{Star}$ ≈ 0.30±0.01 $R_{Sun}$ for ages ranging all the way from 150 Myr to 10 Gyr. Similarly, assuming a reasonable age of ≈ 5 Gyr suggested by its kinematic properties, intermediate between young disk and old disk, i.e., between ≈ 3 and 10 Gyr (Delfosse et al. 1998), the radius predicted by these tracks is 0.30 $R_{Sun}$ (consistent with the 0.29 $R_{Sun}$ cited by B05b) and recent measurements by van Braun et al (in press). Combined with the empirical luminosity, the latter radius yields an effective temperature $T_{eff}$ = 3561K; we thus adopt the rounded-off value $T_{eff}$ = 3600K[2].

We point out that, for the inferred mass, the evolutionary tracks themselves predict $T_{eff}$ ≈ 3440K for ages of 150 Myr to 10 Gyr, or ≈ 150K lower than our value. The difference is most likely due to the sub-solar metallicity in Gl581. Bonfils et al. (2005a) found [M/H] = -0.25 for this star; while Johnson & Apps (2009) have now shown that the metallicities predicted by the former are likely to be underestimations, they still find Gl581 to be slightly metal-poor, with [M/H] = -0.1.

For a given age, lowering the metallicity is predicted to have hardly any effect on radius, but significantly increase effective temperature, and thus the luminosity (Chabrier & Baraffe 2000). Thus the radius we derive from the solar-metallicity tracks is expected to be accurate, but the $T_{eff}$ inferred by combining this with the empirical luminosity should indeed be higher than in these tracks if Gl581 is somewhat metal-poor.

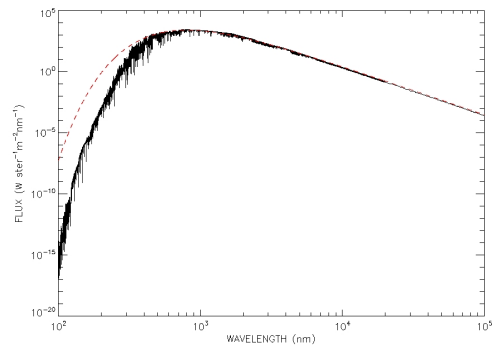

**Fig.1: Flux of Gl581**

We note that recent papers (Selsis et al. 2007, von Bloh et al. 2007, vanParis et al. 2010, Wordsworth et al. 2010) assume $T_{eff}$ = 3200K, much lower than our adopted value, the measured value (van Braun et al. in press) and that predicted by the Baraffe tracks. The reasons for doing so are not apparent, and it forces them to adopt a radius of 0.38 $R_{Sun}$ to remain consistent with the luminosity. This is 30% larger than the tracks predict for a reasonable age; Gl581 would have to be remarkably young (of order ≈ 50 ± 100 Myr according to the tracks) for this radius to be applicable (and if it were this young, its observed luminosity would be 30% smaller than predicted by the tracks). Instead, the range $T_{eff}$ ≈ 3450-3600K, defined by the solar-metallicity track prediction at the lower end and our calculations on the upper, is reasonable, with the higher value preferred given the slightly depressed metallicity in Gl581.

---

[1] Online at http://perso.ens-lyon.fr/isabelle.bara_e/BCAH98 models.

[2] Formally, our rounded-off $T_{eff}$ corresponds to $R_{Star}$ = 0.293 $R_{Sun}$. The slight offset from 0.30 $R_{Sun}$ is of no importance, since radius enters into our analysis only in the temperature determination, and the 40K error in the latter from rounding has negligible impact on the overall stellar flux distribution.

This justifies our adopted value of 3600K. We note that the distance to the Habitable Zone is calculated using only the luminosity of the star, and therefore the location of the HZ inferred by Selsis et al. (2007) is in agreement with ours; however, the difference in our adopted $T_{eff}$ does influence the planetary atmosphere due to the attendant change in the incident stellar spectral energy distribution distiguishing our work from recent models (von Paris et al 2010, Wordsworth et al 2010, von Bloh et al. 2007).

Finally, the magnitude of the stellar chromospheric UV radiation incident on the planet is also very important for calculating the planetary atmospheric properties (Segura et al, 2005). With very scant UV observations available for the vast majority of M dwarfs, chromospheric and coronal H$\alpha$ and X-ray emission have usually been used as proxies for the expected UV emission from these stars. Walkowicz et al. (2008) have shown that some M dwarfs with very low H$\alpha$ and X-ray emission may still have non-negligible near-UV emission, so the use of the former two as a proxy for the latter may not always be accurate. Indeed, there is very likely no such thing as an M dwarf with *no* chromosphere; some basal level of activity is probably always present. However, with absolutely no indication of activity in H$\alpha$ or X-rays in Gl581, our conservative approach is to assume a negligible UV as well, instead of speculating at this juncture. Future direct observations of UV are required to settle the issue. The model stellar flux used is presented in Fig. 1; it is constructed from PHOENIX DUSTY spectra (Allard et al. 2001) to represent a star with $T_{eff}$ = 3600K, log[g] = 5.0, [M/H] = -0.1 and no activity-related emission.

## 3. ATMOSPHERE MODEL DESCRIPTION

Here we use "Super-Earths" as planets that differ from giant planets (Jupiter- and Neptune-like ones) in that they have a surface: a solid-to-gas or liquid-to-gas phase transition at their upper boundary similar to Earth (see also Valencia et al. 2007a, Sasselov et. al. 2008, Sotin et al. 2007, Zahnle et al. 2007). That surface separates a vast interior reservoir (e.g., Earth's mantle) from an atmosphere with insignificant mass compared to that of the planet. Like on Earth, the atmosphere is fed from the interior reservoir, and its chemical balance is achieved via interactions with the interior (outgassing and burial) and with the parent star (photochemistry and loss to space). These interactions are usually described in terms of geochemical cycles; e.g. on Earth the carbonate-silicate cycle maintains a temperate environment through a feedback cycle of outgassing, rain-out and burial of $CO_2$ (see discussion in §5).

For Gl581d we adopt a mass and radius of 7 $M_\oplus$ and 1.69 $R_\oplus$ respectively. We scale the surface gravity according to the increase in mass and radius (see Valencia et al. 2006, 2007, Seager et al. 2007, Sotin et al. 2007). Assuming outgassing and loss rates similar to Earth's, the increase in gravity should proportionally increase the surface pressure. We thus initially set the surface pressure to 2.45 bar in our models, and investigate a range of $CO_2$ levels (implying e.g. an active carbonate-silicate cycle and resulting $CO_2$ buildup to stable condition between outgassing and rainout) to assess habitable conditions on the outer edge of the HZ. We assume efficient heat transfer from the day to the night side in the atmosphere as well as explore the effect of limited heat tranfer and tidal-locking by parametrizing the transported solar flux (for a discussion on the problems to establish any realistic 3D model for Gl581d see Wordsworth et al. 2010a, Wordsworth et al. 2010b).

In our models we focus on two scenarios for rocky super-Earth atmospheres, [Model A] assume a total surface pressure of 2.45bar while [Models B] use a surface pressure of 7.6bar in accordance to minimal conditions for habitability (§4.1). We model atmospheric composition from high oxygen and low $CO_2$ content (A1) to high $CO_2$ and low oxygen content (A1, B1, B2, B3). These two atmospheric compositions should roughly bracket conditions due to an active carbonate-silicate cycle on a rocky planet in the outer part of the HZ.

To model the atmospheric composition, temperature and spectral features, we use Exo-P, a coupled one-dimensional code developed for rocky exoplanets based on the 1D climate (Kasting et al. 1984a, 1984b, Haqq-Misra et al. 2008), 1D photochemistry (Kasting & Ackerman 1986, Pavlov et al. 2000, Kharecha et al. 2005, Segura et al. 2007) and radiative transfer model based on SAO98 (Traub & Stier 1978, Kaltenegger & Traub 2009) – to self consistently calculate the atmosphere and hypothetical spectra of Gl581d, assuming a rocky Earth-analog composition. These codes have been used to calculate HZs around different types of host stars (Kasting et al. 1993) as well as model spectra for Earth-analogs around different host stars (Segura et al. 2003, 2005, 2007; Kaltenegger & Traub 2009, Kaltenegger & Sasselov 2010) and throughout geological evolution of Earth (Kaltenegger et al. 2007).

### 3.1 Coupled 1-D atmosphere models

We used two versions of the 1-D atmosphere models in order to simulate atmospheres with two different compositions: oxygen rich present Earth analog and high $CO_2$ atmospheres. The model for high $O_2$ and low $CO_2$ atmospheres is described by Segura et al. (2005). For $CO_2$ rich atmospheres we coupled the photochemical model used in Segura et al. (2007) with a modified version of the radiative/convective model (Haqq-Misra et al. 2008).

Both versions of the photochemical model used grids with constant altitude (1 km for the high $O_2$ model and 0.5 for the high $CO_2$ model), and solve the continuity equation at each height for each of the long-lived species, including transport by eddy and molecular diffusion. The combined equations were cast in centered finite difference form. Boundary conditions for each species were applied at the top and bottom of the model atmosphere, and the resulting set of coupled differential equations was integrated to steady state using the reverse Euler method. The high $CO_2$ photochemical model has 73 chemical species involved in 359 reactions, while the high $O_2$ version solves for 55 different chemical species that are linked by 217 separate reactions. Photolysis rates for various gas-phase species were calculated using a δ two-stream routine (Toon et al., 1989) that accounts for multiple scattering by atmospheric gases and by sulfate aerosols, for the high $CO_2$ model hydrocarbon aerosols are included too. One important feature of the high $CO_2$ model is its ability to keep track of the atmospheric hydrogen budget, or redox budget. H and $H_2$ escape was simulated by assuming an upward flux at the diffusion limited rate (Walker, 1977). The temperature profile used in the photochemical model is that calculated for the radiative/convective model.

The radiative/convective model incorporates a δ 2-stream scattering algorithm (Toon et al., 1989) to calculate the visible/near IR fluxes and uses four-term, correlated-$k$ coefficients to parameterize absorption by $O_3$, $CO_2$, $H_2O$, $O_2$, and $CH_4$ in each of 38 spectral intervals (Kasting and Ackerman, 1986). The mid infrared is calculated by two different procedures. Ozone, oxygen methane and stratospheric water concentrations used by this model are calculated by the photochemical model. For low $CO_2$ atmosphere the radiative/convective model uses the rapid radiative transfer model (RRTM) algorithm (Mlawer et al. 1997). The high $CO_2$ version of this model uses the hemispheric mean two-stream approximation (Toon et al. 1989).

For both, high and low-$CO_2$ atmosphere, a fixed relative humidity was assumed following Manabe and Wetherald (1967). The tropospheric lapse rate follows a moist adiabat (Kasting, 1988) that takes into account $CO_2$ and $H_2O$ condensation. For all the models $N_2$ concentration is calculated to fill out the atmosphere after the concentrations of the other chemical species have been set up.

## 3.2 Radiative transfer model (SAO model)

The spectral line database of the radiative transfer code includes the HITRAN 2008 compilation plus improvements from pre-release material and other sources (Rothman et al. 2004, 2009, Yung & DeMore 1999). The far wings of pressure-broadened lines can be non-Lorentzian at around 1,000 times the line width and beyond; therefore, in some cases ($H_2O$, $CO_2$, $N_2$), we replace line-by-line calculation with measured continuum data in these regions. The detailed line-by-line modeling required to accurately describe the transfer of radiation in planetary atmospheres is currently restricted to 1-D simulations. Rayleigh scattering is approximated by applying empirical wavelength power laws (Allen 1976, Cox 2000). We assume that the light paths through the atmosphere can be approximated by incident parallel rays, bent by refraction as they pass through the atmosphere, and either transmitted or lost from the beam by single scattering or absorption. We model the spectra using its spectroscopically most significant molecules, $H_2O$, $O_3$, $O_2$, $CH_4$, $CO_2$, CO, $H_2S$, $SO_2$, OH, $H_2O_2$, $H_2CO$, O, $HO_2$, $CH_3Cl$, HOCl, ClO, HCl, $NO_2$, NO, $HNO_3$, $N_2O_5$, $N_2O$ and $N_2$, where $N_2$ is included for its role as a Rayleigh scattering species and broadening gas. We do line-by-line radiative transfer through the refracting layers of the atmosphere.

## 3.3 Gl581d atmosphere simulations

In our models we focus on two scenarios for rocky super-Earth atmospheres, [Models A] assume a nominal surface pressure of 2.45bar while [Models B] are set to a surface pressure of 7.6 bar that allows for habitable conditions in our models (§4.1).

We initially set the surface pressure to 2.45 bar in our models in accordance with the increase in gravity. [Model A1] assumes present Earth-like atmospheric composition with 0.21 $O_2$ and 335 ppmv or 1PAL of $CO_2$ (PAL = Present Atmospheric Level)).We then increase the amount of $CO_2$ from 1 PAL (A1) to a mixing ratio of 0.9 (A2) to explore if the surface temperature of the planet would rise above freezing without a surface pressure increase.

Models B explore the minimum amount of $CO_2$ needed to maintain an average surface temperature above freezing on the planet's surface in an atmosphere with a 0.9 $CO_2$ mixing ratio with either abiotic (B1) or two different biotic levels of oxygen (B2 and B3). The main characteristics of each atmosphere are presented in Table 1.

### 3.3.1 Earth-analog Model atmosphere

Boundary conditions for the biotic oxygen rich atmosphere (A1) were set following Segura et al. (2005) for present

Earth-like atmospheres around quiescent M dwarfs: fixed velocity deposition for $H_2$ of $2.4\times10^{-4}$ cm/s, fixed velocity deposition for CO of $1.2\times10^{-4}$ cm/s, fixed flux for $N_2O$ of $1.131\times10^9$ cm$^{-2}$s$^{-1}$, fixed flux of $CH_3Cl$ of $4.117\times10^8$ cm$^{-2}$s$^{-1}$, and fixed mixing ratio for $CH_4$ of $5\times10^{-4}$.

### 3.3.2 High $CO_2$ Model atmospheres

For the high $CO_2$ simulated atmospheres (A2 and B1-3) the methane surface flux is set to present Earth's abiotically produced $CH_4$ surface flux of $4.13\times10^9$ molecules s$^{-1}$ cm$^{-2}$ ($1.8\times10^{13}$ g yr$^{-1}$, see discussion on Segura et al 2005) that represents ~3% of the total $CH_4$ present Earth production of $1.24\times10^{11}$ molecules s$^{-1}$ cm$^{-2}$ ($5.35\times10^{14}$ g yr$^{-1}$, Houghton et al. 1995). $CO_2$ concentration is fixed in our simulation, consistent with the assumption that the carbonate-silicate cycle controls the $CO_2$ atmospheric abundance on rocky planets (e.g. Kasting and Catling, 2003).

The volcanic outgassing rates in the present Earth adopted for this purpose were $2\times10^{10}$ cm$^{-2}$ s$^{-1}$ for $H_2$ (Holland 2002) and $3.5\times10^9$ cm$^{-2}$ s$^{-1}$ for $SO_2$ (Kasting 1990). The high $CO_2$ version of the photochemical model was modified to account for the potentially higher volcanic activity in super-Earths resulting from larger internal heat, by including three times the present Earth's volcanic outgassing of hydrogen ($H_2$) and $SO_2$. The eddy diffusion profile was scaled from the one measured on present Earth (Massie & Hunten 1981). The oxygen content in the high $CO_2$ atmospheres was modeled in two different ways: an abiotic scenario with a fixed deposition velocity of $O_2$ (Models A2 and B1), and a biotic scenario with two fixed mixing ratio of $O_2$ (Models B2 and B3). In the first case the sources of $O_2$ are chemical reactions in the atmosphere and water photolysis (see Segura et al. 2007). Higher concentrations of $O_2$ require the presence of biological activity, as modeled in the second scenario, where we use initial $O_2$ mixing ratios of $10^{-3}$ (B2) or $10^{-2}$ (B3) for a fixed surface pressure of 7.6 bar (see discussion).

## 4. RESULTS

In §4.1 we present the results for our various model atmospheres, focusing in particular on the surface temperatures they imply, assuming rapid rotation or 100% efficient heat transfer. We then present, in §4.2, the expected spectra for our minimum-case habitable scenario. The detectability of the salient spectral features, in light of the star-to-planet contrast ratio, is discussed in §4.3. The constraints set by 1:1 tidal-locking are subsequently explored in §4.4.

### 4.1 Model Atmospheres

We begin with our results for models A, where the total surface pressure is scaled to 2.45 bar to account for the higher gravity on Gl581d relative to Earth. We examine two cases within this scenario: the first is an Earth-analog oxygen-rich model (A1), which has biotically produced $O_2$ with a terrestrial mixing ratio of 0.21 and a $CO_2$ mixing ratio equal to 1 PAL. We assume a $CH_4$ ougassing of $4.13\times10^9$ molecules s$^{-1}$ cm$^{-2}$ (see Sec. 3.1), that scaled to Gl581d surface is $4.93\times10^9$ g yr$^{-1}$.

Table 2 presents the main result for model A1: the expected surface temperature as a function of radial distance, over the range 0.12 AU (the 1 AU equivalent orbit around Gl581) to 0.22 AU (slightly beyond the 0.20 AU orbital radius of Gl581d). We see that the average surface temperature for an Earth-like atmosphere falls below freezing for radii $\gtrsim$ 0.14 AU. Thus, an Earth-like super-Earth with a 2.45 bar atmosphere cannot maintain a mean surface temperature above freezing at Gl581d's position (see also §4.4). Fig. 2 shows the atmospheric temperature and mixing ratio profiles as a function of distance from the star for 1 PAL and 10 PAL $CO_2$

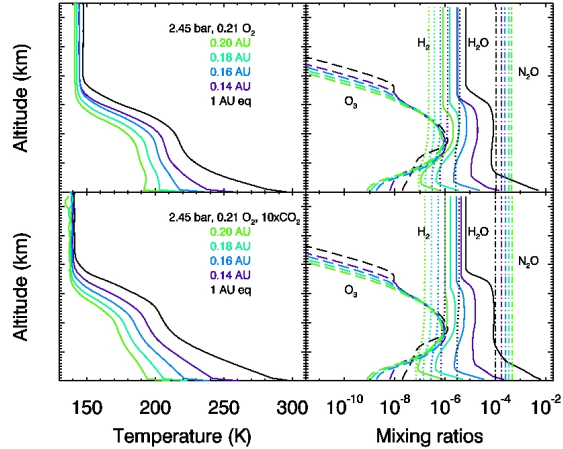

Fig. 2: (left) Temperature profile for a rapid rotating Gl581d-like planet as a function of distance from its host star in the HZ for an Earth-like atmosphere (A1) (upper panel) and 10 PAL CO2 (lower panel). (right) Mixing ratios versus height for the major atmospheric gases. Methane and oxygen are not plotted because they have constant mixing ratios on these atmospheres (see text).

Next, we raise the atmospheric $CO_2$ level in our models to assess what level is actually required in the planet to maintain its mean surface temperature above freezing, and thus provide a habitable environment at its orbit. This in turn sets the surface pressure on the planet. Fig. 3 shows that even a 90% $CO_2$ (and low, abiotic oxygen) planetary atmosphere yields a temperature of only 237 K – well below freezing – at Gl581d's 0.2 AU orbital radius, if the total surface pressure remains fixed at 2.45 bar (model A2).

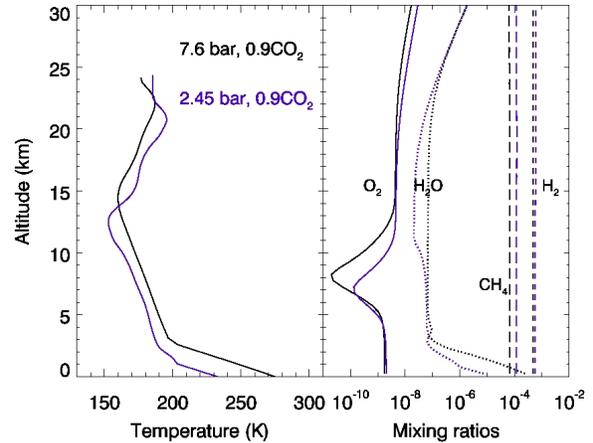

Fig. 3: Temperature and mixing ratios for a 2.45 bar (A2) versus a 7.6 bar (B1) abiotic high $CO_2$ atmosphere.

However, increasing the surface pressure to 7.6 bar while maintaining this high level of $CO_2$ (model B1) increases the surface temperature to an above-freezing value of 275 K (Fig. 3), implying habitable conditions. Such a level of atmospheric pressure increase due to $CO_2$ is reasonable with even a moderate carbonate-silicate cycle or increased outgassing of $CO_2$.

Note that methane production in these models is kept to abiotic levels (see section 3.1), which results in $CH_4$ concentrations of $1.21\times10^{-4}$ and $1.13\times10^{-4}$ in models [A2] and [B1] respectively. While increasing methane levels can also amplify the greenhouse effect on a planet, we concentrate here on $CO_2$ levels supplied by the carbonate-silicate cycle to define the outer edge of the HZ, which encapsulates our assumption that it is the carbonate-silicate cycle that regulates habitability.

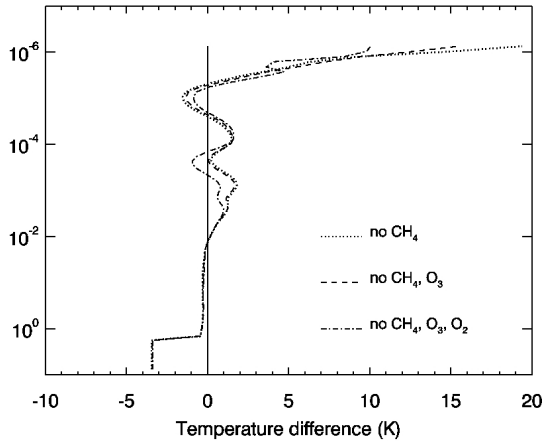

Fig. 4: Temperature difference versus pressure when removing individual component successively from the abiotic high $CO_2$ atmosphere (B1) model for Gl581d.

We use model [B1] as a representative atmosphere for minimal condiations for potential habitability on Gl581d to explore the influence of the individual chemicals on the temperature structure of the planet's atmosphere. Fig. 4 shows the influence of the individual main chemicals on the atmospheric temperature profile in detail. T=0 corresponds to the temperature of the B1 atmosphere model. The x-axis in fig. 4 shows the temperature difference that corresponds to removing individual chemicals ($CH_4$, $CH_4$ & $O_2$, $CH_4$ & $O_2$ & $O_3$) from the B1 atmosphere model. Removing $CH_4$ from the B1 atmosphere lowers the surface temperature as well as the temperature in the upper atmosphere, while warming the intermediate height. Removing $O_3$ slightly cools the upper atmosphere, while removing removing $O_2$ also cools the intermediate atmosphere as well as the upper atmosphere. Model [B1] includes only abiotic sources of oxygen (see Sec. 3), that yields $O_2$ concentrations between $10^{-8}$ and $10^{-9}$ for the 7.6 bar atmosphere.

Having found [B1] to yield habitable conditions, we now explore how much *biotic* oxygen such an atmosphere can support. We examine two such models, both for a 7.6 bar atmosphere: [B2], with an $O_2$ mixing ratio of $10^{-3}$ and 90% $CO_2$; and [B3], with an $O_2$ mixing ratio of $10^{-2}$ and 90% $CO_2$. We adopt model [B2] (7.6 bar atmospheric pressure, 90% $CO_2$ and $10^{-3}$ $O_2$), which provides above-freezing surface conditions for a minimum atmospheric pressure and $CO_2$ mixing ratio, as our habitable biotic scenario for Gl581d to assess remotely detectable spectral features (§4.2, §4.3).

### 4.2 Detectable Spectral Features

In this section we focus on the detectable atmospheric spectral features of a potentially habitable Gl581d. We concentrate on the case of efficient heat transfer, following recent studies demonstrating that heat transfer should not be limited on a planet with a high density atmosphere even in a synchronous rotation (see Joshi et al. 2003; Edson et al. 2011. Wordsworth et al. 2011b, Heng et al. 2011). In particular, we use as our template the atmospheric model [B2] (7.6 bar atmosphere, with 90% $CO_2$ and biotic $O_2$ with a mixing ratio of $10^{-3}$), which is habitable both in the absence of tidal-locking as well as under conditions of relatively efficient heat-transfer in the presence of tidal-locking (§4.4). For simplicity, we restrict ourselves here to spectral features arising from this model in the absence of tidal-locking. This is justified (§4.4) to the extent that the planet is unlikely to be synchronous rotating (see e.g. Leconte et al. 2010), or even if it were, the heat transfer should be efficient, given that at least 77% of its surface is likely directly illuminated (see discussion in Selsis et al. 2007).

The spectral distribution of M-stars, such as Gl581, differs from that of the Sun in two important ways. Their lower effective temperatures shifts the peak of the spectral energy distribution redwards, decreasing the photospheric flux in the ultraviolet (this is mitigated to the extent that ultraviolet emission can arise from the intense chromospheric activity observed in a large fraction of M dwarfs; however, given that there are no signs of such activity in Gl581, as discussed in §2, we do not consider this possibility any further here). This in turn generates a different photochemistry on planets orbiting within the HZ of M stars, compared to planets within the HZ of Sun-like stars (Segura et al. 2005). In particular, the biogenic gases $CH_4$, $N_2O$, and $CH_3Cl$ have substantially longer lifetimes and higher mixing ratios than on Earth, making them potentially observable by space-based telescopes. In addition, the low effective temperatures of M dwarfs yield spectra dominated by molecular absorption bands that redistribute the radiated energy in a distinctly non-black-body fashion. Both effects are crucial to determining the observable spectral features and biosignatures of habitable planets (see e.g. Kaltenegger et al. 2010a) around these cool low-mass stars. We discuss the resulting emergent spectra in §4.3.1, and the transmission spectra in §4.3.2.

### 4.2.1. Emergent Spectra

Fig. 5 shows the spectra and detectable biosignatures in the atmosphere of Gl581d in emergent spectra. The lower panels show the influence of the individual chemicals on the overall spectrum. Each chemical has individual absorption features that are indicated. The summed effects of the chemicals produce the overall spectral appearance of the atmosphere (upper panel). Note that multiple scattering in the atmosphere will affect the visible spectrum, what has not been included in our models. In addition cloud coverage would reduce the depth of the spectral features further (see Kaltenegger et al. 2007, Fujii et al. 2011).

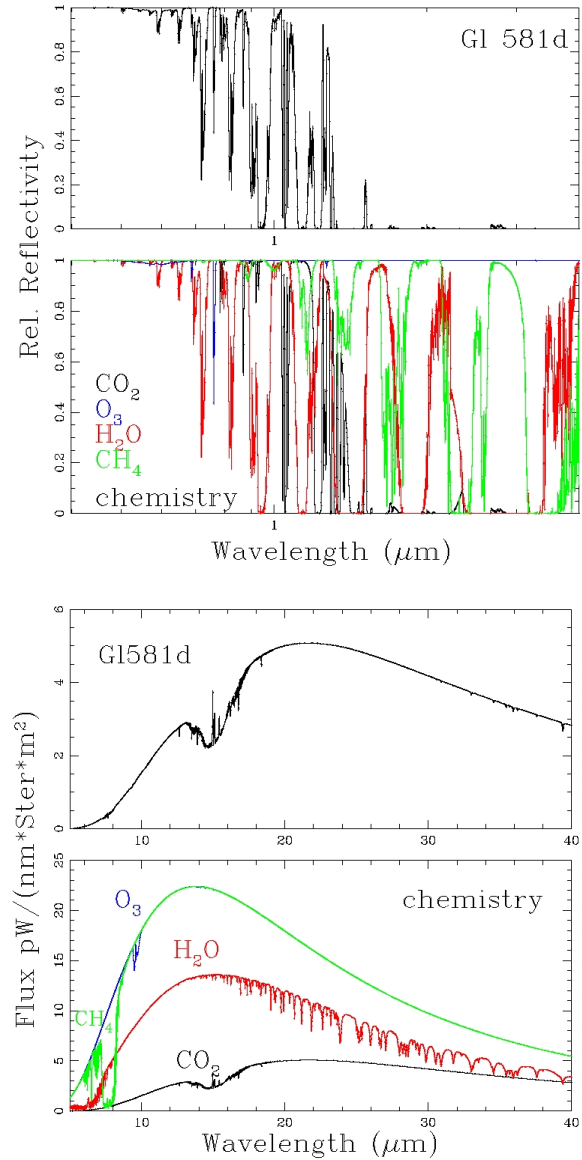

**Fig. 5: Detectable spectral features and biosignatures in the emergent spectra of Gl581d over 0.4-20 microns, for the [B2] model atmosphere. Left: VIS to NIR (0.4-4 μm). Right: MIR (4-40μm).**

Fig. 5 demonstrates that in a dense atmosphere individual chemicals can dominate and blanket all other atmospheric features (see also Selsis 2002). The extremely strong $CO_2$ features dominate the spectra from 1.2 μm onwards. In the mid-infrared (> 4μm), the dense $CO_2$ atmosphere also affects the overall flux emitted by the planet by lowering the effective temperature (see Fig. 5). The visible is not sensitive to high $CO_2$ concentrations, and is not influenced by the atmospheric absorption of $CO_2$, assuming a clear atmosphere. Features of water and oxygen can be seen in the visible.

### 4.2.2 Transmission Spectra

We calculate the transmission spectra for Gl581d. Even though there is no indication yet that Gl581d is a transiting planet, these features could be detected by a space mission like JWST if either Gl581d or a similar planet transits (see Kaltenegger & Traub 2009 for details on geometry).

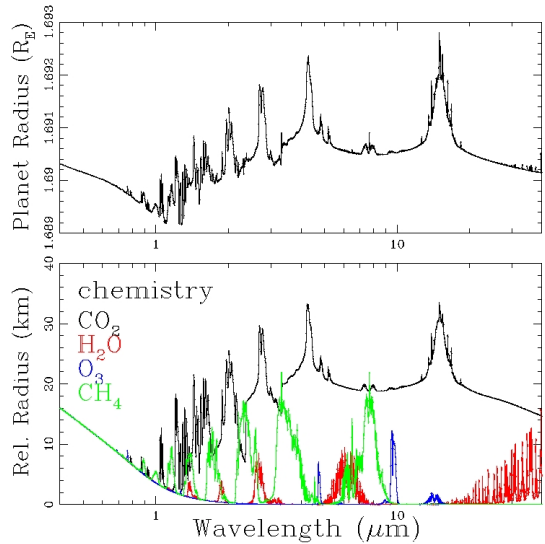

**Fig. 6: Detectable spectral features and biomarkers in the transmission spectra of Gl581d in the VIS to mid IR 0.4-40 μm (left). The overall spectrum is shown on the upper panel, the individual chemicals in the lower panel (B2 model atmosphere).**

Fig. 6 shows the model spectrum of Gl581d in transmission. Even though $CO_2$, $H_2O$, $O_3$ and $CH_4$ are easily visible in the panel that shows individual absorption by chemicals, the overall spectrum is dominated by $CO_2$ down to 1 μm and Rayleigh scattering below that wavelength, not providing information about the habitability of a rocky planet with a dense $CO_2$ atmosphere. Cloud coverage would reduce the depth of the spectral features further (see §5).

### 4.3 Planet Star Contrast Ratio

The observable quantity to remotely derive what atmospheric features exist in a planet's atmosphere is the contrast ratio versus wavelength, as shown in Fig. 7 and Fig. 8 for emergent and transmission model spectra respectively. The Sun emits a large fraction of its energy in the visible, a wavelength domain where the atmosphere of a habitable planet is highly reflective, because of the Rayleigh backscattering varying like $\lambda^{-4}$ and because of the lack of strong $H_2O$ absorption bands. The emission of Gl581, a star with a low effective temperature, peaks in the near-infrared where the contribution of Rayleigh scattering to the albedo becomes negligible and the strong absorption bands of $H_2O$, $CO_2$ and $CH_4$ cause additional absorption of stellar radiation and overall lower the planet's albedo as long as no additional reflective cloud layer forms (see §5).

For the emergent spectra, the larger surface area of a Super-Earth makes the direct detection and secondary-eclipse detection of its atmospheric features and biosignatures easier than for Earth size planets. Even though Gl581d orbits its host star at a distance less than 1AU, the integrated flux received by the planet is half the flux received by Earth. Since Gl581d orbits an M star, it has a lower Bond albedo and thus a smaller part of the stellar light is reflected, making Gl581d appear dimmer in reflected light than an equivalent planet orbiting at 1AU around a sun-like star. Even though the bond albedo decreases from Earth's 0.29 to Gl581d's 0.12, the increased surface area increases the overall reflected flux of the planet by about 15% over Earth's reflected Sunlight (see Fig. 7). In the MIR the flux increases proportional to the surface area of the planet if the effective temperature is equal. In our model the decrease in effective temperature for Gl581d due to the high $CO_2$ concentration, which reduces its flux compared to the Earth significantly.

the planetary atmosphere reduces the detectable absorption features of a rocky planet with a dense atmosphere like Gl581d in transmission compared to a planet with Earth's gravity (see Fig. 8). Note that the detection of the planet itself is easier in transmission because of its increased size.

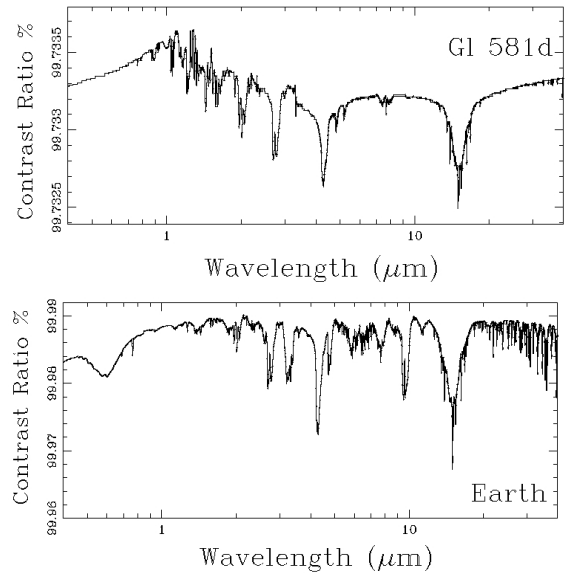

**Fig. 8: Planet-star contrast ratio for transmission spectra from 0.4 to 40 μm for a clear atmosphere (left) B2 model atmosphere (right) Earth (adapted from Kaltenegger & Traub 2009).**

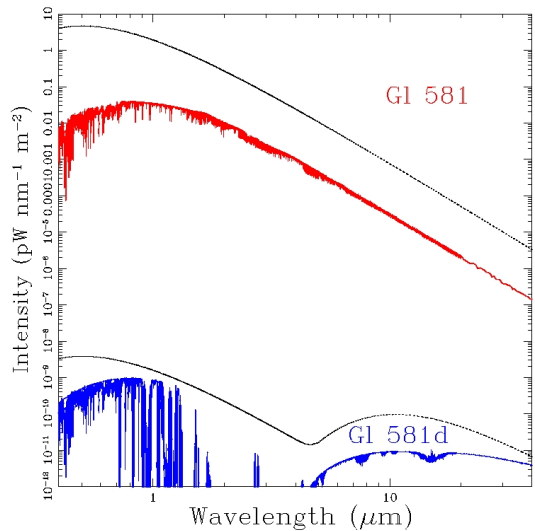

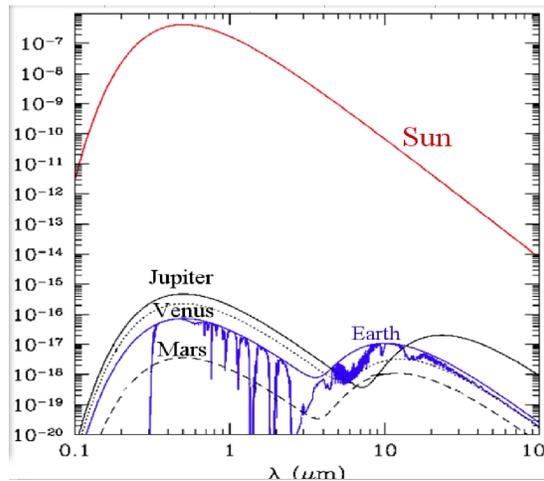

**Fig. 7: Planet-star contrast ratio for emergent spectra (assuming 1/2 illumination of the planet) from 0.4 to 40 μm for a clear atmosphere (left) B2 model atmosphere (green indicated the Earth-Sun system level) ), (right) Sun and Earth (adapted from Traub & Jucks 2002).**

### 4.4. Exploring the effect of synchronous rotation

For an M dwarf, the HZ is so close to the star that the planets are likely to become tidally locked within a relatively short time after they form (Dole, 1964; Kasting et al., 1993). Tidally locked planets can be locked into several resonances, especially if other planets are present, as in the case of the Gl581 system (see e.g. Leconte et al. 2010, Heller et al. 2011, Barnes et al. 2009, Henning et al. 2009). For any resonance except synchronous rotation, all parts of planetary atmosphere are illuminated and there is no permanently dark area (see also Selsis et al. 2007). Detailed models have shown that even for planet's in synchronous rotation direct illumination of

The increase of the gravity and resulting decrease in the atmospheric scale height of only one hemisphere does not prevent habitability, for planets with even modestly dense atmospheres (Haberle et al. 1996, Joshi et al. 1997, Joshi 2003, Wordsworth et al. 2010a, Edson et al. 2010, Heng et al. 2011). Especially if an ocean is present, even a tidally locked Earth should remain habitable, provided atmospheric cycles transport heat from the dayside to the nightside.

We explore the effects of tidal locking or limited heat transfer in the atmosphere in two ways: 1) For model atmospheres that are found to have surface temperatures close to, but not above, freezing for the non-tidally-locked case (i.e., for even surface illumination), we determine whether tidal-locking can push the system into habitable conditions by raising dayside temperatures above freezing while avoiding permanent $CO_2$ condensation on the nightside. 2) For model atmospheres which are found to already have surface temperatures above freezing (i.e., be habitable) in the absence of tidal-locking, we explore whether locking can hinder habitability by causing permanent $CO_2$ condensation on the nightside. The partial pressure of water in all our models is below the condensation limit at all altitudes on the night side of the simulated atmospheres.

For the extreme case of synchronous rotating planets with permanent night sides, the lower temperature on the dark side needs to be calculated to see if chemical species like $H_2O$ and $CO_2$ permanently condense out, thereby destabilizing the entire atmosphere and in effect moving the outer edge of the HZ inwards for a synchronous rotating planet. The fraction of the planet's atmosphere that would have to be above a certain temperature to induce water evaporation and its later photo-dissociation and loss of hydrogen to space at the inner edge, and the permanent condensation of $CO_2$ at the outer edge of the HZ, is unknown and will require detailed dynamical 3D modelling for such planets (see Wordsworth et al. 2010a, 2010b on the problems generating a 3D model for Gl581d). Assuming efficient heat transfer in the atmosphere, the limits of the HZ are unlikely to change significantly due to synchronous rotation.

In the absence of 3D models, we parameterize the heat transfer on the planet to explore the effect of slow and synchronous rotation. Due to the dense atmosphere and in accordance with preliminary findings of 3D models (Wordsworth et al. 2010a) we explore a maximum difference of 20% in overall stellar heat flux being deposited on the day and night side, what translates into an average surface temperature difference of 40K (see also Wordsworth et al. 2010a). This corresponds to a parameterization of 10% more and 10% less stellar incident flux respectively on the top of the planetary atmosphere. In terms of remotely detectable features, this mainly influences the infrared emission spectrum of the planet, since geometry dictates that one will always see the terminator region between day and night side for transiting planets and no flux from the nightside for the reflection spectrum.

To explore the effect of synchronous rotation in our atmospheric models, we assume that one hemisphere of Gl581d receives direct starlight over some part of the orbit while the other is in perpetual darkness (this is a conservative estimate, since a larger fraction of the planetary surface is expected to be directly illuminated; see discussion in Selsis et al. 2007). Further we assume that this produces an uneven distribution of heat flux over the two hemispheres due to inefficient atmospheric heat transfer, and calculate the resulting increase and decrease in the average hemispheric surface temperatures.

Because no *photons* are available over the dark hemisphere except for scattered light in the atmosphere, the chemistry on the dark side of the atmosphere will change depending on the efficiency of material transport from the dayside. For this, detailed 3D dynamical calculations are needed (see also Edson et al. 2011). As a first order approximation we calculated the residence times for the most abundant compounds after $CO_2$ and $N_2$: $O_2$, $H_2$, $CH_4$, and CO. These lifetimes were calculated considering that the photolysis source of each compound was removed, that is the column depth ($cm^{-2}$) of a compound divided by the sum of all the integrated production rates ($cm^{-2} s^{-1}$) of the photolysis reactions that result on that compound. Following Joshi et al. (1997) we calculated the advective time scale with winds of 10 m/s (terrestrial value) for a planet with our calculated radius for Gl581d that resulted to be about 40 days. This means that the atmosphere will be transported to the night side on that time scale. On Joshi et al (1997) the simulation with a 1 bar atmosphere shows a strong superrotation with winds between 10-110 m/s. According to these authors' results, larger surface pressures will result on smaller temperature difference between the day and the night side associated to more vigorous circulation in the atmosphere. For Gl 581d we would expect winds strong enough to keep the advective time scale shorter than the lifetimes of the main compounds of the simulated atmospheres, that means that these chemical species are likely to be in the atmosphere long enough to be transported to the night side.

Note that, in addition to the overall gradient in temperature between the two hemispheres due to this inefficiency in heat transfer, there will also in general be regions of higher and lower temperature over a given hemisphere, compared to the hemispheric mean. Here we are mainly interested in the change in the mean hemispheric temperature caused by synchronous rotation, and the implications for habitability.

We first examine the two models, [A2] and [B3], whose surface temperatures were found to be below freezing, but not greatly so, in the non-tidal locking situation (§4.1). Our goal is to see whether tidal-locking can make at least their dayside hemisphere habitable. In other words, we examine how inefficient the heat transfer would have to be to raise surface temperatures above freezing on the dayside, and whether this inefficiency would in the end preclude habitability by causing permanent $CO_2$ condensation on the nightside due to the correspondingly low temperatures there.

For model [A2] (2.45 bar atmosphere, with 90% $CO_2$ and abiotic $O_2$), we find that the effective heat flux on the dayside hemisphere has to be 40% higher than obtained for even heat distribution, to raise its temperature above freezing. In other words, out of the total stellar heat flux received by the planet (all of which is incident on the dayside, by definition), 70% must *remain* on the dayside, and just 30% can be transferred by the atmosphere to the nightside. But a heat flux equaling only 30% of the total would cause a freeze-out of $CO_2$ on the nightside, thereby destabilizing the entire atmosphere. We thus conclude that tidal-locking cannot induce habitability for this particular model. This result reinforces the need, noted earlier in §4.1, for higher surface pressures on Gl581d to provide habitable conditions (see Fig. 9).

Conversely, we may enquire whether this small inefficiency in heat transfer can drive model [B2] (7.6 bar atmosphere, with 90% $CO_2$ and biotic $O_2$ with a mixing ratio of $10^{-3}$), which was found to be habitable under non-tidally-locked conditions (§4.1), to *un*inhabitable surface temperatures. Fig. 10 shows that this is not the case: the dayside surface temperature, for a heat flux increase of 10% over that obtained with even heat distribution, changes from 275 K to 292 K (still consistent with liquid water), while the corresponding decrease in the nightside heat flux reduces the temperature there to 256 K (still sufficiently high to prevent a freeze-out of $CO_2$). This temperature difference corresponds to preliminary finding o 3D models (Wordsworth et al 2010a). Thus, Gl581d would continue to be habitable with this model atmosphere in spite of tidal locking. A similar trend is seen in model atmosphere [B3].

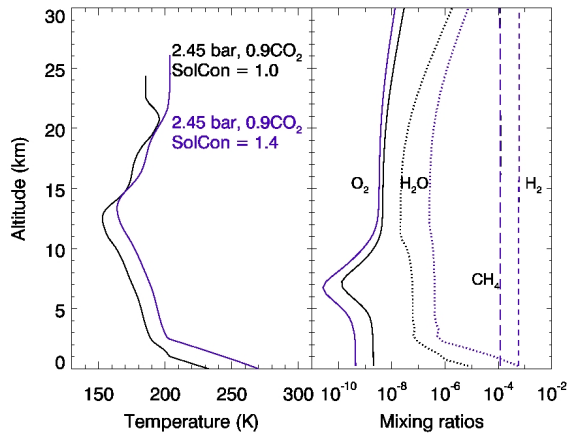

Fig. 9: (left) Temperature profile for the dayside of Gl581d for a rapid rotating (solid line) as well as tidally locked planet (dashed line) that has an average surface temp above freezing (40% more flux on the dayside) for an abiotic high $CO_2$ 2.45bar atmosphere (A2). (right) Mixing ratios versus height for the main atmospheric gases.

In conclusion, both the models with high atmospheric pressure and biotic oxygen that we've considered, [B2] and [B3], are likely to be habitable under the small heat transfer inefficiencies likely to result from tidal-locking.

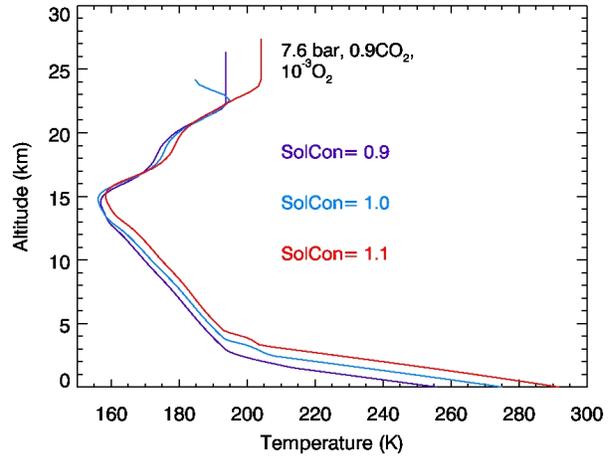

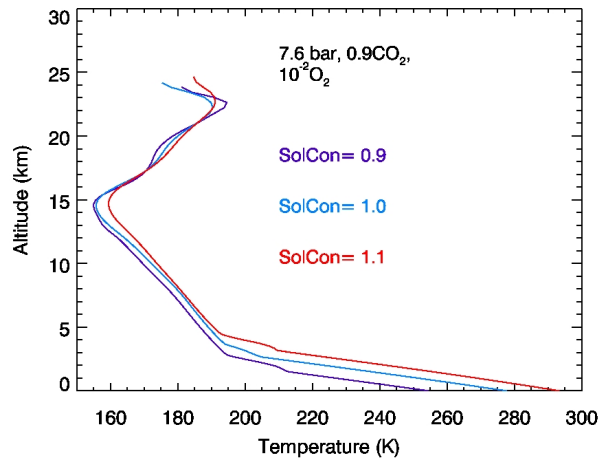

Fig.10: Temperature profile for the dayside and nightside of Gl581d for a rapid rotating (solid line) as well as tidally locked planet (dashed lines) with 10% more flux on the dayside than on the nightside for the biotic 7.6 bar (left) low oxygen atmosphere (B2) (right) (B3).

## 5. DISCUSSION

### 5.1 The Effect of $CO_2$ Clouds

We assume here that the planet is rocky and is dominated by a global carbonate-silicate cycle that stabilizes the surface temperature and the $CO_2$ level, like on Earth (Walker et al. 1981), according to models for super-Earths (Valencia et al. 2007b, Sotin et al. 2007, Zahnle et al. 2007). This implies that the planet is geologically active, continuously outgasses $CO_2$ and that carbonates form in the presence of surface liquid water, which may require continental weathering. Without this stabilization the Earth would not be habitable. This in turn affects the composition of the atmosphere at the outer edge of the Habitable Zone: The amount of $CO_2$ increases, the colder surface temperature due to decreased rain out from the atmosphere of the planet is, what in turn increases the greenhouse effect, warming the surface. The feedback only works up to the temperature where the increase in the albedo due to $CO_2$ concentration is stronger than the greenhouse effect and $CO_2$ freezes out in the atmosphere. The warming effect of $CO_2$-ice clouds has not been included in the models. Forget and Pierrehumbert (1997) showed that the warming effect caused by the backscattering of the infrared surface emission can exceed the cooling effect caused by the increase of albedo. Because of their size and their optical constants, typical $CO_2$-ice particles are relatively transparent to visible radiation, but scatter efficiently at mid-infrared wavelengths. For a stellar flux shifted towards the near-infrared, the albedo of these clouds could be significantly higher than for a solar irradiation. As a consequence, $CO_2$ condensation increases the greenhouse warming compared to a purely gaseous $CO_2$ atmosphere (see Wordsworth et al 2010). This makes our calculation a conservative model of the atmosphere temperature: the planet may be potentially habitable for lower values of $CO_2$ than shown here. Depending on the fractional cloud cover, the theoretical outer edge of the HZ for our Sun occurs between 1.67 AU in the cloud-free limit and 2.4 AU for 100% cloud cover (Forget & Pierrehumbert 1997; Mischna et al. 2000). Several groups are currently modeling $CO_2$ clouds to estimate their effect on a planetary climate (see e.g. and Wordsworth et al. 2010, Kitzmann et al. 2010, Goldblatt & Zahnle 2011). The optical properties of $CO_2$-ice clouds have not been included in our calculations and models because no conclusive models are available at this point.

## 5.2. Levels of biotic oxygen in Earth-like models for Gl581d

Note that the model we present here only represents one possible nature of a planet like Gl581d in a wide parameter space that includes Mini-Neptunes. Composition of the planet will influence the atmosphere substantially, e.g. the abundance of oxygen found in the abiotic case strongly depends on the oxydation state of the superficial layers and possible oceans of the planet. If e.g. the surface of the planet were highly oxidized, a higher buildup of oxygen would be expected for the same production levels (see e.g. Zahnle et al. 2007, Kasting et al. 1984b). This parameter space is very wide and we concentrate on Earth analog models here, to explore the case if an Earth-like planet could produce signatures of habitability that we could remotely detect.

## 5.3 Habitability within the HZ

For mostly dry super-Earths in the habitable zones of their stars, namely in an equilibrium temperature range that allows liquid surface water, the geochemical cycles will involve a hydrological component. The carbon cycle, e.g., on early pre-biotic Earth, would be fed by the volcanic outgassing of $CO_2$ which is balanced by the burial of calcium carbonate through silicate weathering reactions that remove protons and increases alkalinity of the oceans (Walker, Hays, & Kasting 1981). The model spectral signatures of a global carbon cycle on the emergent spectrum of a planet over geological time have been computed by Kaltenegger et al. (2007). If e.g. a sulfur cycle were present on a planet, the outer edge of the HZ would shift outwards due to increased greenhouse effect of $SO_2$ in addition to $CO_2$ (see Kaltenegger & Sasselov 2009, Domagal-Goldman et al. in press, ), which would reduce the amount of $CO_2$ needed to maintain Gl581d's surface temperature above freezing. Note that a planet found in the HZ is not necessary habitable, since many factors may prevent surface habitability like the lack of water or ingredients necessary for the emergence of life.

## 6. Conclusion

We show that Gl581d is potentially habitable, assuming the carbonate-silicate cycle controls the atmosphere of the planet. We calculate the surface temperature and atmosphere including potential biomarkers – assuming different atmospheres compositions, high oxygen atmosphere analogous to Earth's as well as high $CO_2$ atmospheres with and without biotic oxygen concentrations (Fig. 2 to Fig. 4). We find that a minimum $CO_2$ partial pressure of about 7 bar, in an atmosphere with a total surface pressure of 7.6 bar, are needed to maintain a mean surface temperature above freezing on Gl581d. The model we present here only represents one possible nature – an Earth-like composition - of a planet like Gl581d in a wide parameter space. The surface temperature of a simulated 90% $CO_2$ and low oxygen planetary atmospheres at 0.2 AU changes from 237 K to 278 K when increasing the surface pressure from 2.45 bar to 7.6 bar surface pressure (see Fig. 3). Such a level of atmospheric pressure increase due to $CO_2$ is reasonable even with a moderate carbonate-silicate cycle or increased outgassing of $CO_2$. Additional greenhouse gases like $CH_4$ and $SO_2$ as well as clouds assuming a net warming could decrease this amount due to their added warming effect.

We model synthetic spectra from 0.4μm to 40μm and show where indicators of biological activities in such a planet's atmosphere could be observed by future ground- and space-based telescopes (Fig. 5: secondary eclipse/direct imaging; Fig. 6: transmission). The model transmission spectrum of Gl581d is dominated by $CO_2$ down to 1 μm and Rayleigh scattering below that wavelength, not providing information about the habitability of a rocky planet with a dense $CO_2$ atmosphere. For the emergent spectra, the larger surface area of a Super-Earth makes the direct detection and secondary-eclipse detection of its atmospheric features and biosignatures easier than for Earth size planets. In the infrared region of the emergent spectrum, $CO_2$ also dominates the atmospheric features (Fig. 5). In the visible part of the emergent spectrum, biomarkers could be detected even for high $CO_2$ concentrations (Fig. 5). Note that we assumed a clear atmosphere and did not include multiple scattering in the atmosphere. Cloud coverage would further reduce the depth of all spectral features.

Our concept of the habitable zone is based on the carbonate-silicate cycle. Even so the measurements are hard (as shown in Fig. 7 and Fig. 8) this concept can be probed by observing detectable atmospheric features by future ground and space- based telescopes like E-ELT and JWST. Observation of the emergent spectrum could also determine if Gl581d is the first habitable world we have discovered.

**Acknowledgements**

L.K. acknowledges support from NAI and DFG funding ENP Ka 3142/1-1. A.S. and L.K. thank the support by project PAPIIT-IN119709–3.

**Table 1. Characteristics of the simulated atmospheres for Gl581d.**

| Model | Mixing ratios | | Surface conditions | | Solar constant (SolCon) | Boundary conditions |
|---|---|---|---|---|---|---|
| | $O_2$ | $CO_2$ | Pressure (bar) | Temp (K) | | |
| A1 | 0.21 (biotic) | $3.35 \times 10^{-4}$ (1PAL) $3.35 \times 10^{-3}$ (10 PAL) | 2.45 | 210 | 1.0 | Constant surface flux of $N_2O$, $CH_3Cl$, and $CH_4$, constant deposition velocity CO and $H_2$. Fixed mixing ratio for $O_2$ and $CO_2$. |
| A2 | Not fixed, abiotically produced | 0.9 | 2.45 | 232 | 1.0 | Constant mixing ratio of $CO_2$. Constant velocity deposition for $H_2$, CO, $O_2$ and $O_3$. Constant surface flux for $CH_4$. |
| | | | | 270 | 1.4 | |
| B1 | Not fixed, abiotically produced | 0.9 | 7.6 | 256 | 0.9 | Same as A2 |
| | | | | 275 | 1.0 | |
| | | | | 290 | 1.1 | |
| B2 | $10^{-3}$ (biotic) | 0.9 | 7.6 | 256 | 0.9 | Constant mixing ratio of $CO_2$ and $O_2$. Constant velocity deposition for $H_2$, CO, and $O_3$. Constant surface flux for $CH_4$. |
| | | | | 275 | 1.0 | |
| | | | | 292 | 1.1 | |
| B3 | $10^{-2}$ (biotic) | 0.9 | 7.6 | 255 | 0.9 | Same as B2 |
| | | | | 278 | 1.0 | |
| | | | | 293 | 1.1 | |

**Table 2. Surface temperature of the 2.45 bar high oxygen planetary atmospheres of Gl581d (A1: 0.21 $O_2$, $5 \cdot 10^{-4}$ $CH_4$, 1PAL $CO_2$ = 335 ppm and 10PAL $CO_2$ = 3350 ppm) vs. distance to its host star from the 1AU equivalent (0.12 AU) to 0.22 AU.**

| D (AU) | Surface parameters $CO_2$ | |
|---|---|---|
| | T (1 PAL) | T (10PAL) |
| 0.12 | 297 | 299 |
| 0.14 | 257 | 260 |
| 0.16 | 237 | 241 |
| 0.18 | 222 | 227 |
| 0.20[a] | 210 | 215 |
| 0.22 | 201 | 206 |

[a] Gl581d's eccentric orbit with a = 0.22 AU corresponds to a circular orbit with a = 0.2 AU.